\documentclass[preprint,aps,12pt,notitlepage,nofootinbib,tightenlines]{revtex4}
\usepackage{amsmath}
\usepackage{bm}
\usepackage{times}
\usepackage{braket}
\usepackage{color}
\usepackage{epsfig}
\usepackage{slashed}
\usepackage{hyperref}
\usepackage{multirow}
\usepackage{booktabs}
\usepackage{array}
\usepackage{float}
\newcommand{\beq}{\begin{eqnarray}}
\newcommand{\eeq}{\end{eqnarray}}
\newcommand{\be}{\begin{equation}\begin{aligned}}
\newcommand{\ee}{\end{aligned}\end{equation}}

\newcommand{\gev}{\text{GeV}}

\definecolor{Red}{rgb}{1.,0.,0.}

\definecolor{Blue}{rgb}{0.,0.,1.}

\definecolor{nicered}{rgb}{0.7,0.1,0.1}
\definecolor{nicegreen}{rgb}{0.1,0.5,0.1}
\def\lsim{ {\ \lower-1.2pt\vbox{\hbox{\rlap{$<$}\lower6pt\vbox{\hbox{$\sim$}}}}\ } }
\def\gsim{ {\ \lower-1.2pt\vbox{\hbox{\rlap{$>$}\lower6pt\vbox{\hbox{$\sim$}}}}\ } }

\hypersetup{colorlinks,citecolor=nicegreen,linkcolor=nicered}
\begin{document}
\title{Single production of vectorlike $T$ quark at future high-energy linear $e^{+}e^{-}$ collider}
\author{Lin Han$^{1}$, Liu-Feng Du$^{2}$, Yao-Bei Liu$^{2}$\footnote{E-mail: liuyaobei@hist.edu.cn}}
\affiliation{1. School of Biomedical Engineering, Xinxiang Medical University, Xinxiang  453003, China\\
2. Henan Institute of Science and Technology, Xinxiang 453003, China}

\begin{abstract}
Based on a model-independent framework including the vectorlike top partner (VLQ-$T$),
we investigate the prospect of discovering the singlet or doublet VLQ-$T$ via the single production process $e^{-}e^{+}\to T\bar{t}+t\bar{T}$ with the $T\to Zt$ decay channels at a future high-energy linear $e^{+}e^{-}$ collider with $\sqrt{s}=3$~TeV. We focus on the hadronic
 decay of the top quark and two types of decay channel for the $Z$ boson: $Z\to \ell^{+}\ell^{-}$  and $Z\to \nu\bar{\nu}$.
By carrying out a full simulation for the signals and the relevant Standard Model backgrounds, the $2\sigma$ exclusion limit and $5\sigma$ discovery prospects are, respectively, obtained on the VLQ-$T$ mass and the coupling strength $g^{\ast}$ with the integrated luminosity of 5~ab$^{-1}$.
In addition, we considered the initial state
radiation and beamstrahlung effects as well as the systematic uncertainty effects of backgrounds, which are found to reduce the excluding or discovery capability.
\end{abstract}

\maketitle
\newpage
\section{Introduction}
To solve the gauge hierarchy problem~\cite{DeSimone:2012fs}, new vectorlike quarks~(VLQs) are introduced to regulate the Higgs boson mass-squared divergence in many new physics models beyond the Standard Model~(SM), such as little Higgs model~\cite{ArkaniHamed:2002qy}, composite Higgs model~\cite{Agashe:2004rs}, and other extended models~\cite{He:1999vp,Wang:2013jwa,He:2001fz,He:2014ora}. A common feature of these VLQs is that the left- and right-handed components transform with the same properties under the SM electroweak symmetry group~\cite{Aguilar-Saavedra:2013qpa}.
Based on the electric charges of $+2/3e$~($T$ quark), $-1/3e$~($B$ quark), $+5/3e$~($X$ quark), or $-4/3e$~($Y$ quark), the VLQs could be grouped in multiplets, such as electroweak singlet [$T$, $ B$], electroweak doublets [ $\left(X,T\right),\left(T,B\right)$ or $\left(B,Y\right)$], or electroweak triplets [$\left(X,T,B\right)$ or $\left(T,B,Y\right)$]. Furthermore, they are expected to couple preferentially to third-generation quarks and can  generate characteristic signatures at the current and future high-energy colliders~(for example see~\cite{Atre:2011ae,Buchkremer:2013bha,Barducci:2017xtw,Cacciapaglia:2018qep,
Liu:2017rjw,Liu:2017sdg,Liu:2019jgp,Tian:2021oey,Yang:2021btv,Moretti:2016gkr,Moretti:2017qby,
Carvalho:2018jkq,Benbrik:2019zdp,Buckley:2020wzk,Brown:2020uwk,Deandrea:2021vje,Belyaev:2021zgq,Dasgupta:2021fzw,Han:2022npb,Cacciapaglia:2021uqh}). Here, we focus on the singlet  or doublet VLQ-$T$ quark, which only couples to third-generation SM quarks.

Using Run 2 data, the direct searches for such VLQ-$T$ have been performed by the ATLAS and CMS Collaborations and the constraints on their masses have been obtained at a 95\% confidence level~(C.L.)~\cite{Aaboud:2018wxv,Aaboud:2018xpj,Aaboud:2018uek,Aaboud:2018ifs,Sirunyan:2018qau,Sirunyan:2018omb,Aaboud:2018pii,CMS:2019eqb}. For instance, the minimum mass of a singlet~(doublet) VLQ-$T$ is set at about 1.31~(1.37) TeV
from direct searches by the ATLAS Collaboration with an integrated luminosity of 36.1 fb$^{-1}$~\cite{Aaboud:2018pii}. The CMS Collaboration have excluded $T$-quark mass below 1.37 TeV at 95\% C.L. by using 35.9 fb$^{-1}$ of $pp$ collision data in the fully hadronic final state~\cite{CMS:2019eqb}.

Due to a much cleaner environment, the future high-energy linear $e^{+}e^{-}$ colliders such as the Compact Linear
Collider (CLIC)~\cite{CLIC1,CLIC2,Franceschini:2019zsg} can probe TeV scale electroweak
charged particles well above the LHC reach.  For instance, the final stage of CLIC operating at an energy of 3 TeV is expected to
directly examine the pair production of new heavy top partners of mass up to 1.5 TeV \cite{Dannheim:2013ypa}.
For the single production process, any such new particle can be produced at CLIC with a sizable rate up to the
kinematic limit of  3 TeV~\cite{Kitano:2002ss,Kong:2007uu,Senol:2011nm,Harigaya:2011yg,Guo:2014piv,Liu:2014pts}.
Furthermore, the single VLQ production could reveal the electroweak properties of the interactions between VLQs and SM particles, which could serve as a complementary
channel to its pair production and thus will be an important task
for future high-energy colliders once the heavy VLQ is discovered at the LHC and its mass determined.

Very recently, the single VLQ production at the future high-energy $e\gamma$ and $e^{+}e^{-}$ colliders was investigated in Reference~\cite{Yang:2018fcx,Shang:2019zhh,Shang:2020clm,Qin2021,Han:2021kcr,Han:2021lpg,Qin:2022mru} with different decay channels. Ref.~\cite{Qin:2022mru}
has investigated the single production of a singlet vectorlike top
partner decaying to $Wb$ at 3 TeV CLIC
and found that the correlation regions of  $g^{\ast}\in [0.15,0.4]$ and $m_T\in$ [1500 GeV, 2600 GeV] can be excluded
with the integrated luminosity of 5 ab$^{-1}$. For a singlet VLQ-$T$, all three decay modes, namely, $T\to Wb$, $T\to tZ$, and $T\to th$,
have sizable branching ratios, while the charged-current decay mode $T\to Wb$ is absent if VLQ-$T$ is either in a
(X,T) doublet.
In this work, we focus on the single production of the singlet or doublet VLQ-$T$ via the decay channel $T\to tZ$ at the future 3 TeV CLIC, which will be a good comparative
study.  We expect that such work  may become complementary to other production processes in searches for the heavy VLQ-$T$ at the future high-energy linear colliders.

This paper is organized as follows: in Sec. II, we brief review  the couplings of VLQ-$T$ with the SM particles  and discuss its single production at the future high-energy $e^{+}e^{-}$ colliders with $\sqrt{s}=3$~TeV.
Section III is devoted to a detailed analysis of the relevant signals and backgrounds. Finally, we give a summary in Sec. IV.

\section{Vectorlike $T$ quark in the simplified model}
\subsection{An effective Lagrangian for vectorlike $T$ quark}
Following the notation of Ref.~\cite{Buchkremer:2013bha}, a generic parametrization of an effective Lagrangian for singlet top quark partners is given by\footnote{Note that the $TWb$ couplings are vanishing for the standard $(T, B)$ doublet VLQ-$T$ quark,  and the model files of the singlet or standard $(T, B)$ doublet VLQ-$T$ can be downloaded from the Feynrules Model Database~\cite{http}.}
\beq
{\cal L}_{\rm eff} =&& \frac{gg^{\ast}}{2\sqrt{2}}[\bar{T}_{L}W_{\mu}^{+}
    \gamma^{\mu} b_{L}+
    \frac{g}{\sqrt{2}c_W}\bar{T}_{L} Z_{\mu} \gamma^{\mu} t_{L}
    - \frac{m_{T}}{\sqrt{2}m_{W}}\bar{T}_{R}ht_{L} -\frac{m_{t}}{\sqrt{2}m_{W}} \bar{T}_{L}ht_{R} ]+ H.c.,
  \label{TsingletVL}
\eeq
where $g$ is the $SU(2)_L$ gauge coupling constant, and $\theta_W$ is the Weinberg angle.  Thus, there are only two model parameters: the VLQ-$T$ quark mass $m_T$ and the coupling strength to SM quarks in units of standard couplings, $g^{\ast}$.

 Certainly, the coupling parameter can also be described as other constants, i.e., $\sin\theta_{L}$~\cite{Aguilar-Saavedra:2013qpa} or $\kappa$~\cite{Buchkremer:2013bha}. After
comparison, we find that there is a simple relation among these coupling parameters:
$g^{\ast}=\sqrt{2}\sin\theta_{L}=\sqrt{2}\kappa$. At 13 TeV LHC,
searches for single production of $T$ quarks have placed limits on $T$-quark production
cross sections for $T$-quark masses between 1 and 2 TeV at 95\% C.L. for various SM couplings~\cite{CMS:2017gsh,CMS:2017voh,CMS:2019afi,ATLAS:2022ozf}. Here, we take a conservative limit for the coupling parameter $g^{\ast}\leq 0.5$, which is consistent with the
current experiment bounds~\cite{ATLAS:2022ozf}. The singlet VLQ-$T$ has three possible decay modes: $T\to$  $bW$, $tZ$, and $th$. For $M_T\geq 1~$TeV, the  branching ratios  are BR$(T\to th)\approx {\rm BR}(T\to tZ)\approx \frac{1}{2}{\rm BR}(T\to Wb)$.  However, the standard $(T, B)$ doublet VLQ-$T$ can decay into $tZ$ or $tH$, each with a branching
fraction of 0.5 in the asymptotic limit where their masses go to infinity, which is a good approximation as expected from the Goldstone
boson equivalence theorem~\cite{He:1992nga,He:1993yd,He:1994br,He:1996rb,He:1996cm}.

\subsection{Single production of VLQ-$T$ at linear  $e^{+}e^{-}$ collider}
\begin{figure}[h]
\centering
\includegraphics[width = 16cm ]{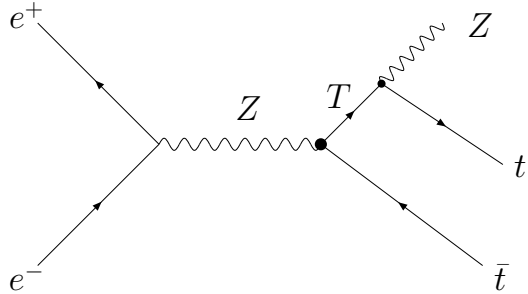}
\vspace{-17cm}
\caption{Representative Feynman diagrams of the processes $e^{+}e^{-}\to T(\to Zt)\bar{t}$.}
\label{fig:fey}
\end{figure}

From the above discussions, we know that the VLQ-$T$  can be singly produced through
$s$-channel $Z$ boson exchange by $e^{+}e^{-}$ collisions.
The relevant Feynman diagrams for single production and decaying into $tZ$ are depicted in Fig.~\ref{fig:fey}.

\begin{figure}[thb]
\begin{center}
\vspace{-0.5cm}
\centerline{\epsfxsize=9cm \epsffile{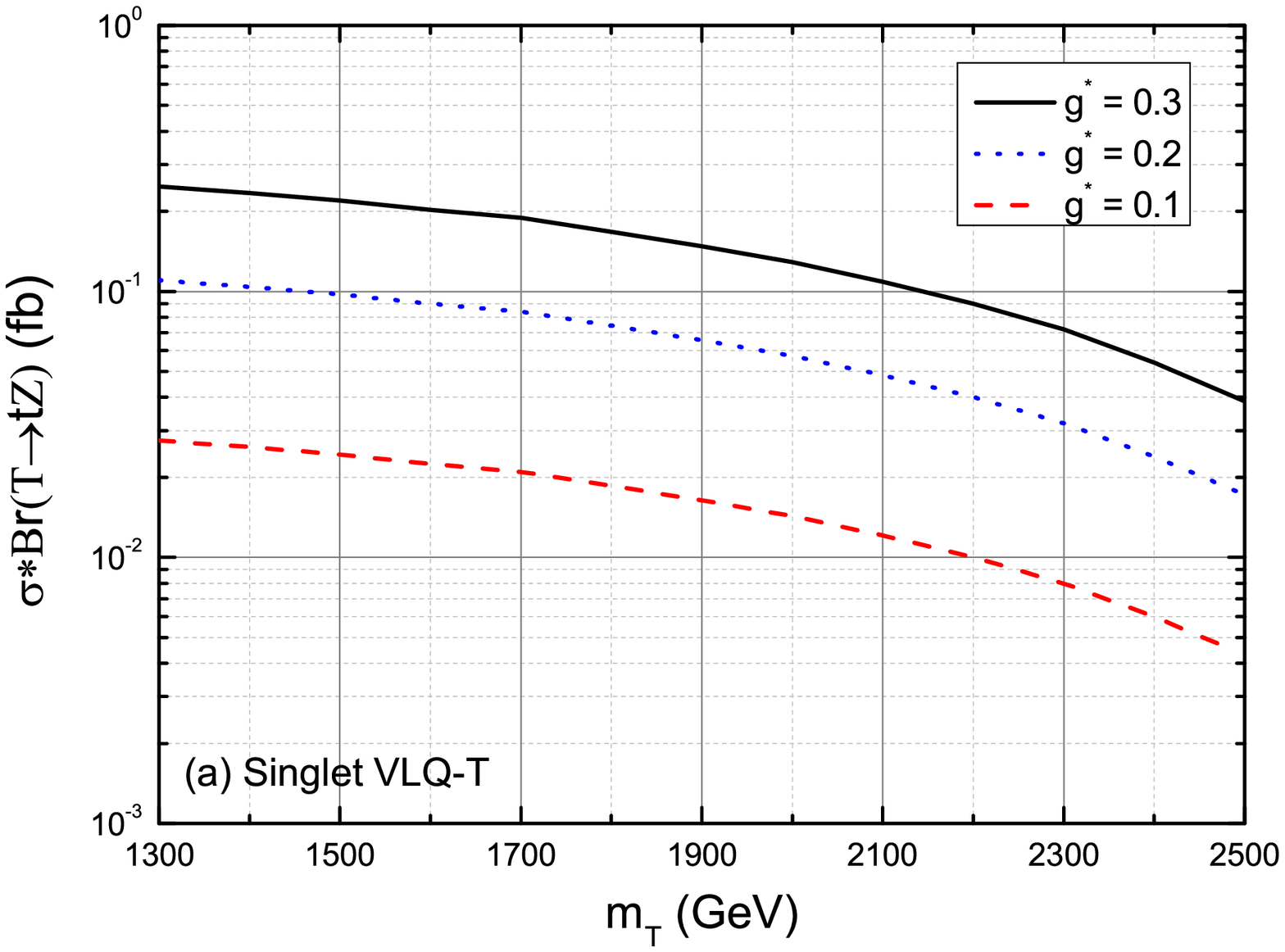}\epsfxsize=9cm \epsffile{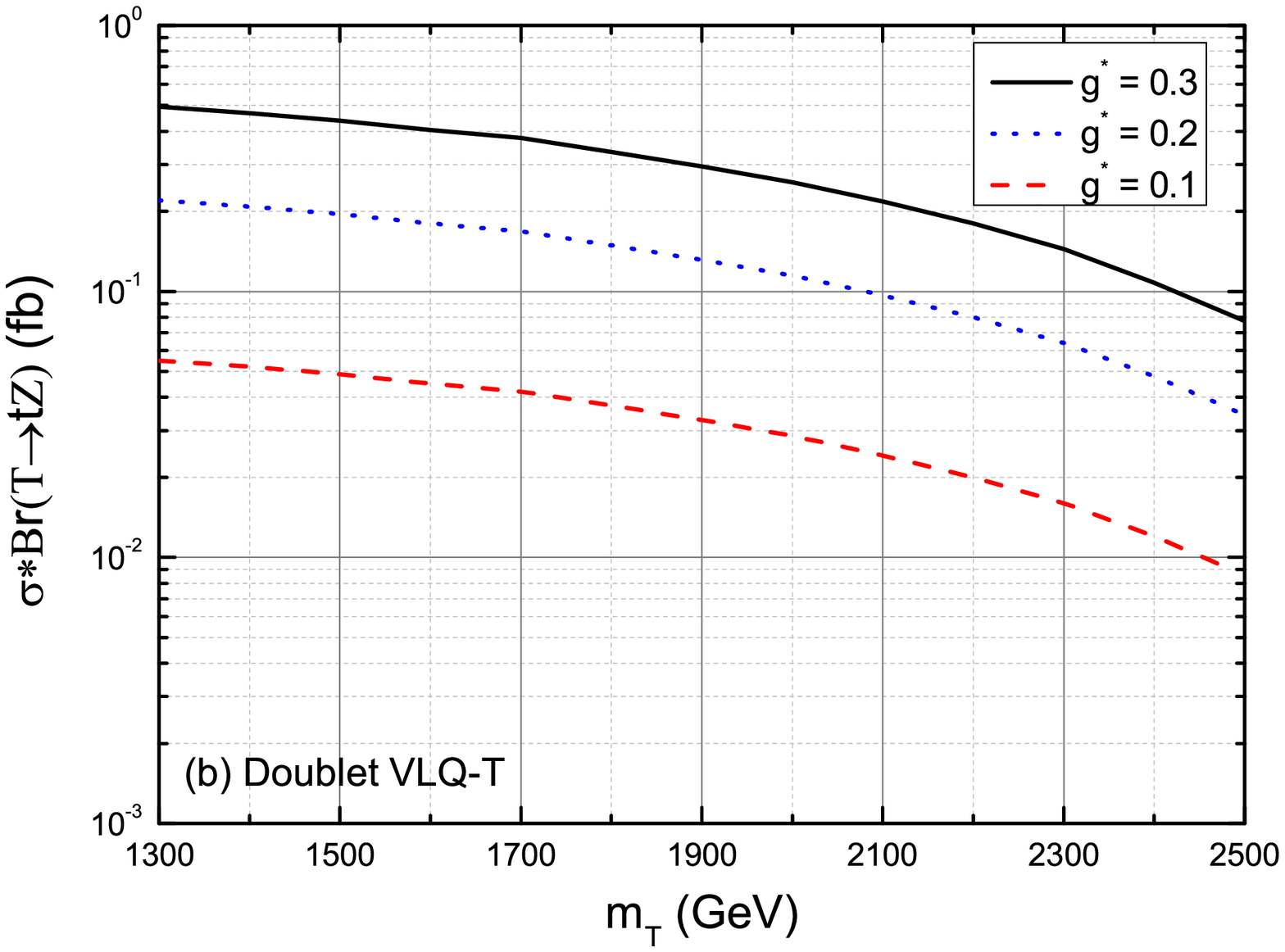}}
\caption{Total cross section of $\sigma\times Br(T\to tZ)$ as a function of $m_T$ with three typical values of $g^{\ast}$ for (a) the singlet case and (b) doublet case. }
\label{cs}
\end{center}
\end{figure}

In Fig.~\ref{cs}, we have shown the dependence of the cross section $\sigma\times Br(T\to tZ)$ for the process $e^{+}e^{-}\to T\bar{t}+t\bar{T}$ without beam polarization on the singlet VLQ-$T$ quark mass
$m_T$ at a 3 TeV CLIC for three typical values of $g^{\ast}$. As the VLQ-$T$ quark mass grows, the cross section of single production decreases slowly due to a larger phase space.
For $g^{\ast}=0.2$ and $m_T=1.5~(2)$ TeV, the cross section can reach 0.1~(0.06) fb.
Obviously, the cross sections of the $(T, B)$ doublet case are about twice as large as the singlet case, and they are proportional to the square of the coupling strength $g^{\ast}$. For the other possible channels, such as $T\to Wb$ and $T\to th$, the cross section $\sigma\times Br(T\to XY)$ can be easily deduced from the relationship between their decay branch ratios, whether for singlet or doublet VLQ-$T$ quarks. For instance, the $(T, B)$ doublet VLQ-$T$ can only decay into $tZ$ or $tH$, with a branching
fraction  $BR(T\to th)\approx {\rm BR}(T\to tZ)\approx50\%$ for a large VLQ-$T$ mass; thus, the cross section $\sigma\times Br(T\to th)$ should be approximately equal to the cross section $\sigma\times Br(T\to tZ)$ for the same parameter values.

\section{Collider simulation and analysis}
Next, we analyze the observation potential by performing a
Monte Carlo simulation of the signal and background events and explore the sensitivity of single VLQ-$T$ via the $T\to tZ$ channel at the 3 TeV CLIC.  Considering the top quark decays hadronically and the subsequent decay channels $Z\to \ell^{+}\ell^{-}$ and $Z\to \nu\bar{\nu}$, respectively, there are two typical final states,
 \begin{itemize}
\item
 $e^{+}e^{-}\to T\bar{t}\to tZ\bar{t}\to bjj\ell^{+}\ell^{-}+X$ for $Z\to \ell^{+}\ell^{-}$ ,
\item
$e^{+}e^{-}\to T\bar{t}\to tZ\bar{t}\to bjj+\slashed E_{T}+X$ for $Z\to \nu\bar{\nu}$.
\end{itemize}
Note that  in order to keep enough signal events, we here do not reconstruct the associated produced top quark and assume that it can decay into anything.

Monte Carlo event simulations for the signal and SM background are generated at leading order~(LO) by using MadGraph5-aMC$@$NLO \cite{mg5}. All event samples are interfaced to Pythia 8.20~\cite{pythia8}  for fragmentation and showering, and then fed into the the Delphes 3.4.2~\cite{deFavereau:2013fsa} for a fast detector simulation, where we choose the  CLIC detector card designed for 3 TeV~\cite{Leogrande:2019qbe}. In our analysis, jets are clustered with the Valencia Linear Collider~(VLC) algorithm~\cite{Boronat:2014hva,Boronat:2016tgd} in exclusive mode. The $b$-tagging efficiency and misidentification rates are taken as the medium working points~(WP) (70\% $b$-tagging efficiency), and the misidentification rates are given as a function of energy and pseudorapidity; i.e., in a bit where $E>500$ GeV and $1.53<|\eta|\leq 2.09$, misidentification rates are $9\times 10^{-3}$ for the medium WP. Finally, event analysis is performed by using MadAnalysis5~\cite{ma5}.

According to above two typical final states, we analyzed the
main backgrounds coming from the SM processes: $t\bar{t}Z$, $WZjj$, $t\bar{t}$, $ZZZ$ and $ZZjj$.
As we know, initial state radiation (ISR) and beamstrahlung will affect the
cross section~\cite{Godbole:2006kr,Dalena:2012fg}, and it is necessary to consider these effects in future high-energy linear colliders. We calculate these effects in the updated version~(v3.3.2) of MadGraph5-aMC$@$NLO and
list these results in Table~\ref{ISR} compared with those cross
sections without these effects.  From Table~\ref{ISR}, we can see
that the cross sections with ISR and beamstrahlung would
be reduced  for the signal process, but enhanced for the SM backgrounds. In the following discussions, we consider the effects without and with the ISR and beamstrahlung
effects, respectively.

\begin{table}[ht!]
\fontsize{12pt}{8pt}\selectfont \caption{Comparison of cross sections~(in fb) without ($\sigma$) and with ($\sigma_{ISR}$) ISR and beamstrahlung effect for $g^{\ast}=0.3$ and two typical VLQ-$T$ quark masses, where ``Ratio" stands for $\sigma_{ISR}/\sigma$. \label{ISR}}
\begin{center}
\newcolumntype{C}[1]{>{\centering\let\newline\\\arraybackslash\hspace{0pt}}m{#1}}
{\renewcommand{\arraystretch}{1.5}
\begin{tabular}{C{1.5cm}| C{2.0cm} |C{2.0cm} |C{2.0cm} |C{1.2cm}C{1.2cm}C{1.2cm}C{1.2cm}C{1.2cm} }
\hline
 \multirow{2}{*}{Process}& \multicolumn{3}{c|}{Signals}&\multicolumn{5}{c}{Backgrounds} \\ \cline{2-9}
&1500 GeV& 2000 GeV& 2500 GeV  & $WZjj$ &$t\bar{t}$&$t\bar{t}Z$& $ZZZ$&$ZZjj$\\   \cline{1-9} \hline
$\sigma$&0.92&0.54&0.17&24.82&19.15&1.66&0.36&0.49\\
$\sigma_{ISR}$&0.87&0.45&0.12&27.4&28.74&1.88&0.41&0.55\\ \hline
Ratio &0.95&0.83&0.70&1.10&1.50&1.13&1.14&1.12
\\
\hline
\end{tabular} }
 \end{center}
 \end{table}

\subsection{Analysis of decay channel $Z\to \ell^{+}\ell^{-}$}
In this subsection, we analyze the signal and background events at 3 TeV CLIC through the $Z\to \ell^{+}\ell^{-}$~($\ell=e,\mu$) decay channel.
For this channel, the typical signal is two opposite-sign and same-flavour~(OSSF) leptons, three jets in which at least one is $b$ tagged. The main SM backgrounds come from the following processes:
 \begin{itemize}
\item
 $e^{+}e^{-}\to t\bar{t}Z$ with $Z\to \ell^{+}\ell^{-}$ and $t\to bW^{+}\to bjj$,
\item
$e^{+}e^{-}\to W^{\pm}Zjj$ with $Z\to \ell^{+}\ell^{-}$ and $W^{\pm}\to jj$,
\item
$e^{+}e^{-}\to ZZZ$ with $Z\to \ell^{+}\ell^{-}$ and  $Z\to q\bar{q}$.
\end{itemize}
Note that the contribution from the  process $e^{+}e^{-}\to ZW^{+}W^{-}$ is also included in the process $e^{+}e^{-}\to W^{\pm}Zjj$ with the $W^{\pm}\to Wjj$ decay.
To identify objects, we choose the basic cuts at parton level for the signals and SM backgrounds as follows:
 \be
p_{T}^{\ell}>~20~\gev,\quad
  p_{T}^{j/b}>~25~\gev,\quad
 |\eta_{\ell/b/j}|<~2.5,\quad
 |\eta_{j}|<~5\\
  \ee
where $p_{T}^{\ell, b, j}$, $|\eta_{\ell/b/j}|$ are the transverse momentum  and pseudorapidity of leptons, $b$ jets, and light jets.
\begin{figure*}[htb]
\begin{center}
\centerline{\hspace{2.0cm}\epsfxsize=9cm\epsffile{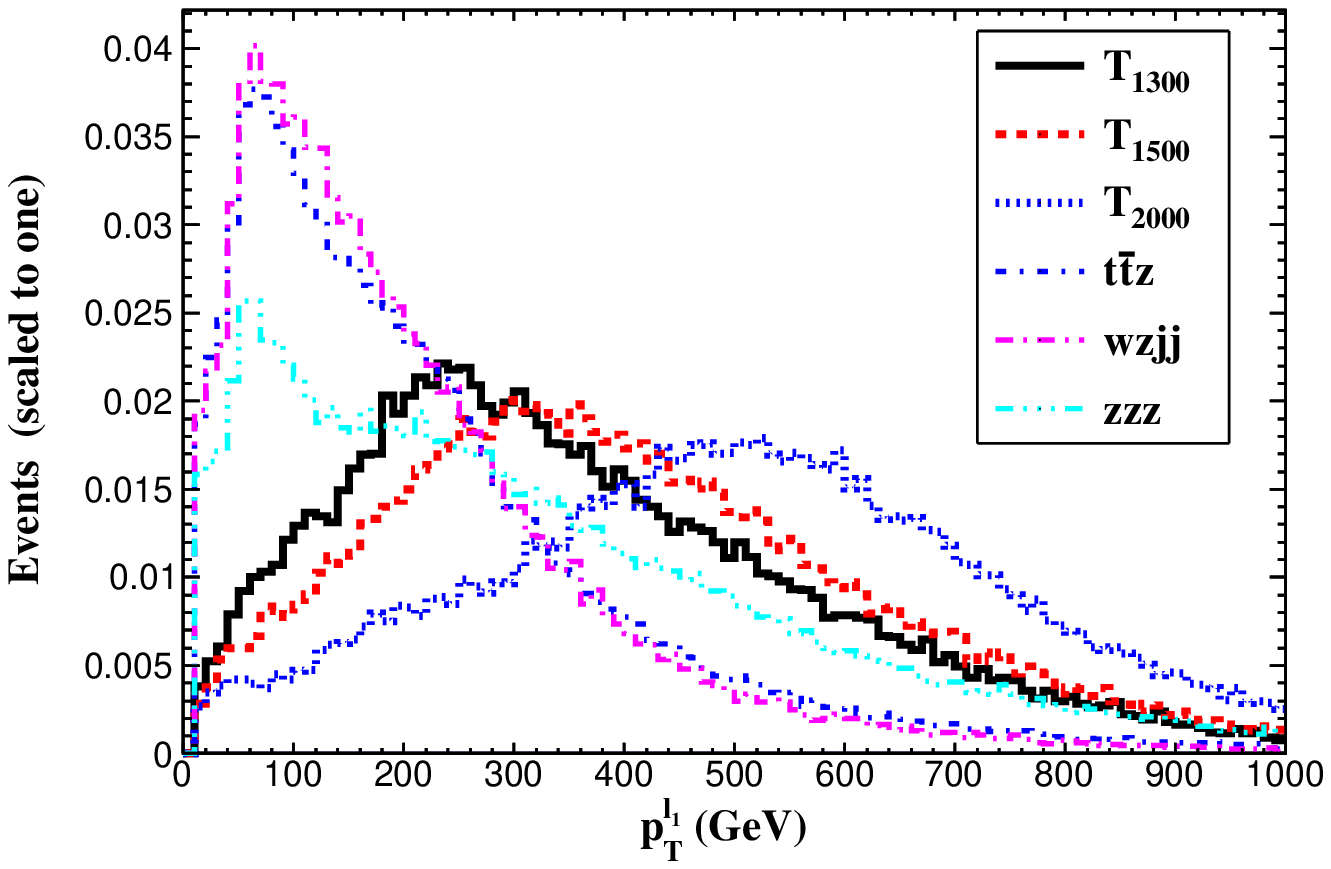}
\hspace{-2.0cm}\epsfxsize=9cm\epsffile{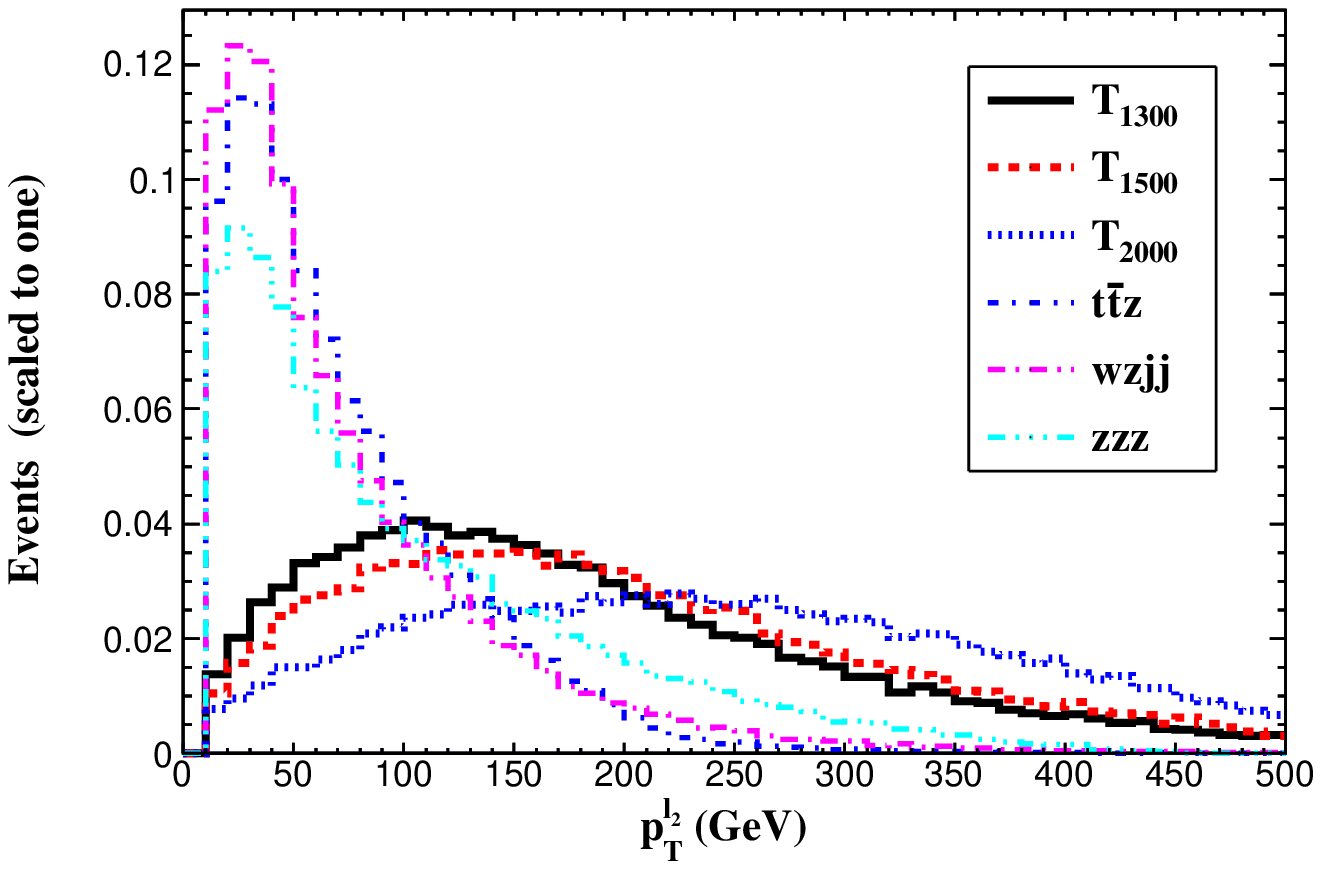}}
\centerline{\hspace{2.0cm}\epsfxsize=9cm\epsffile{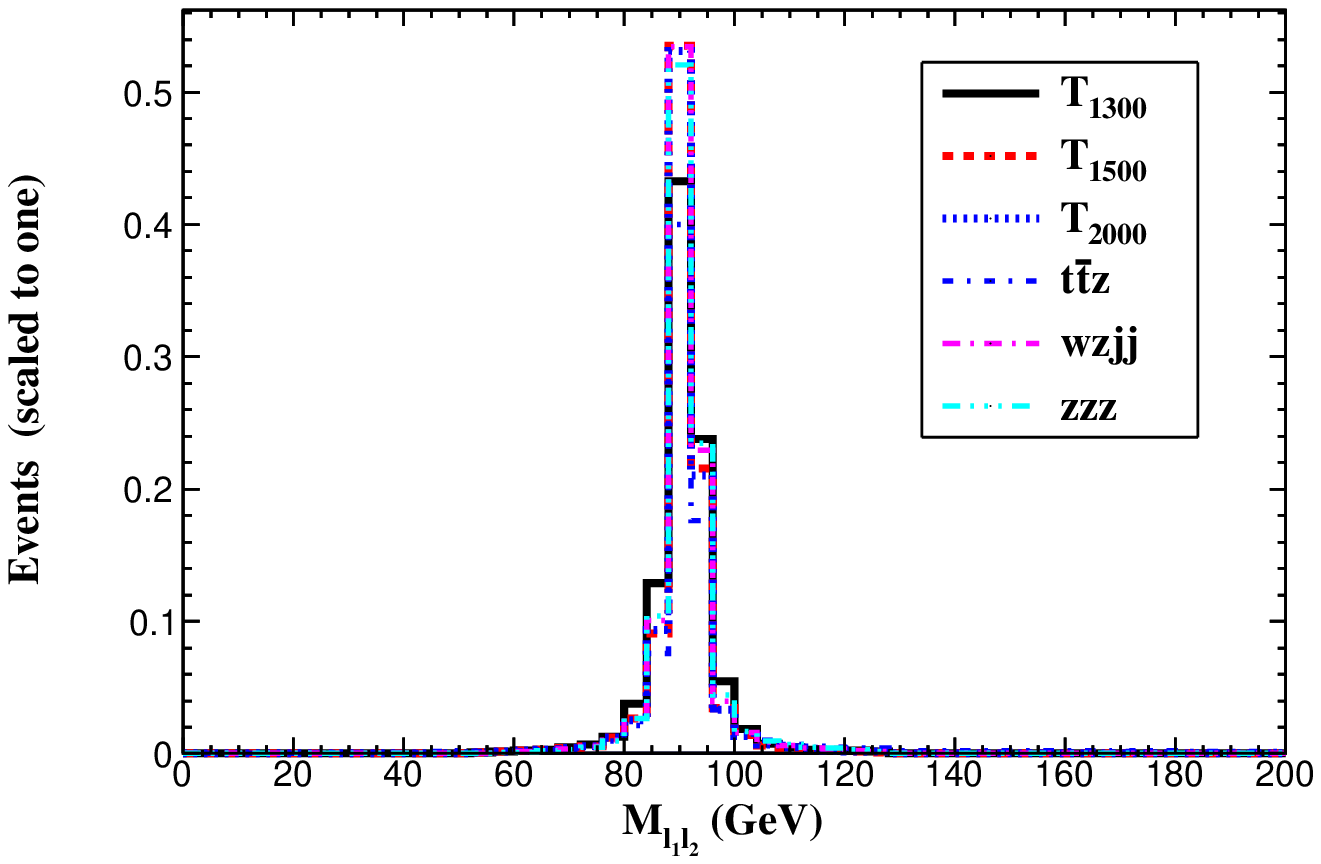}
\hspace{-2.0cm}\epsfxsize=9cm\epsffile{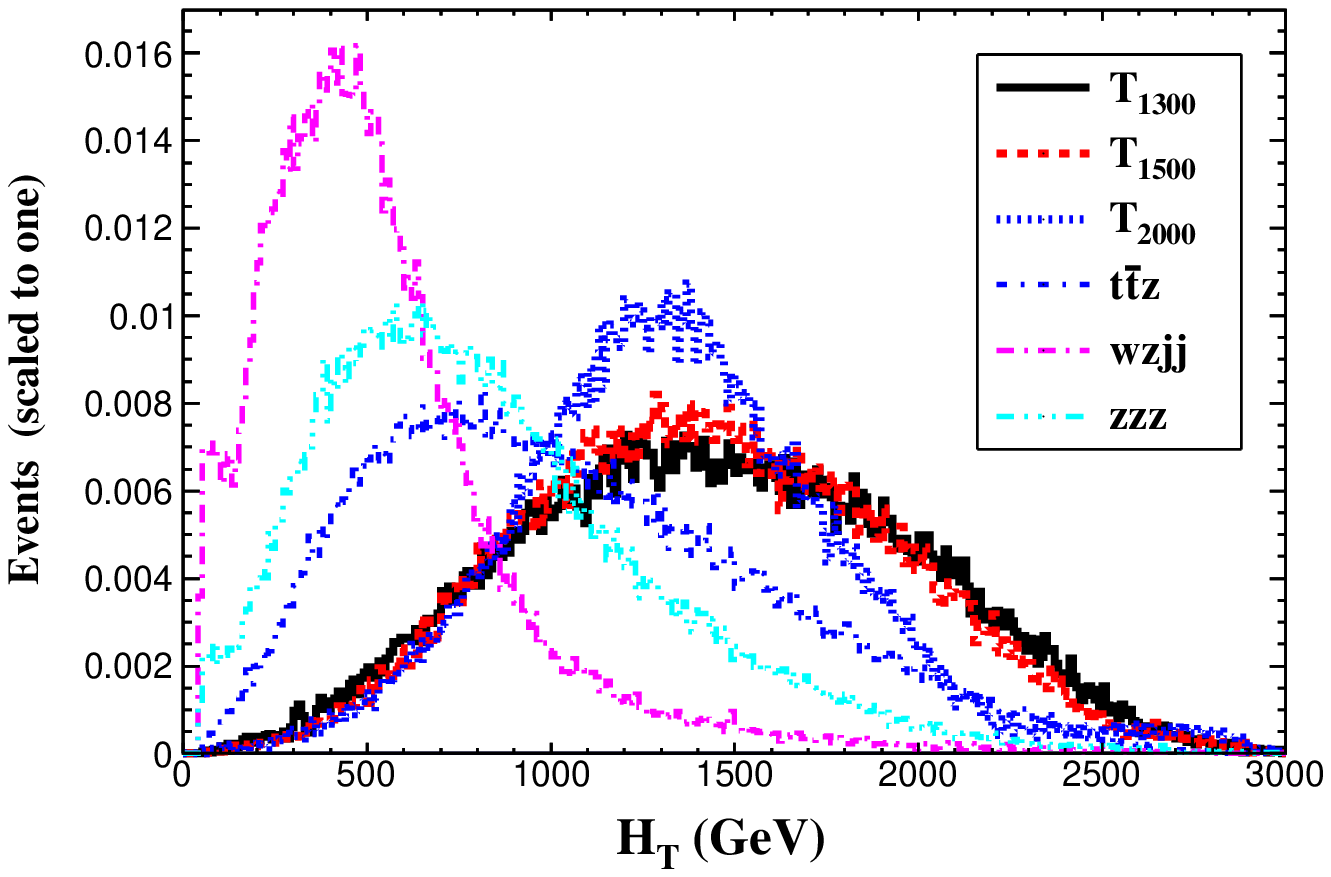}}
\caption{Normalized distributions for the signals (with $m_{T}$=1300, 1500, and 2000 GeV) and SM backgrounds at the CLIC. }
\label{distribution1}
\end{center}
\end{figure*}

For the signal, the leptons $\ell_{1}$ and $\ell_{2}$ are two OSSF leptons that are assumed to be the product of the $Z$-boson decay, and at least three jets are present.
In Fig.~\ref{distribution1}, we plot some differential distributions  for  signals and SM backgrounds, such as the transverse momentum distributions of the leading and subleading leptons~($p_{T}^{\ell_{1}\ell_{2}}$),  the invariant mass distributions of the two leptons~$M_{\ell_{1}\ell_{2}}$, and the scalar sum of the transverse energy of all final-state jets $H_{T}$.
Based on these kinematical distributions, we can impose
 the following set of cuts:
 \begin{itemize}
\item Cut 1: There are exactly two isolated leptons ($N(\ell)=2$), at least three jets in which at least one is $b$ tagged.
\item Cut 2:
     The transverse momenta of the leading and subleading leptons are required to have $p_{T}^{\ell_{1}}> 200 \rm ~GeV$ and $p_{T}^{\ell_{2}}> 100 \rm ~GeV$. Besides, the invariant mass of the $Z$ boson is required to have $|M_{\ell_{1}\ell_{2}}-m_{Z}|< 10 \rm ~GeV$.
\item
Cut 3: The scalar sum of the transverse energy of all final-state jets $H_{T}$ is required to have $H_{T}> 800 \rm ~GeV$.
\end{itemize}

\begin{table}[htb]
\centering %
\caption{Cut flow of the cross sections (in $10^{-3}$ fb) for the signals and SM backgrounds at the 3 TeV CLIC with $g^{\ast}=0.1$ and three typical VLQ-$T$ quark masses. Note that the parenthetical numbers denote the
results with the ISR and beamstrahlung effects.\label{cutflow1}}
\vspace{0.2cm}
\begin{tabular}{p{1.6cm}<{\centering} p{2.0cm}<{\centering} p{2.0cm}<{\centering} p{2.0cm}<{\centering}p{0.3cm}<{\centering}  p{2.0cm}<{\centering} p{2.0cm}<{\centering} p{2.0cm}<{\centering} }
\toprule[1.5pt]
 \multirow{2}{*}{Cuts}& \multicolumn{3}{c}{Signals}&\multicolumn{3}{c}{Backgrounds}  \\ \cline{2-4}  \cline{6-8}
&1300 GeV &1500 GeV& 2000 GeV  && $WZjj$ &$t\bar{t}Z$& $ZZZ$\\    \cline{1-8} \midrule[1pt]
Basic&1.08~(1.078)&0.97~(0.92)&0.57~(0.47)&&357~(393)&20~(22.6)&8.9~(10.15)\\
Cut 1&0.41~(0.409)&0.38~(0.36)&0.23~(0.19)&&25~(27.5)&7.56~(8.54)&1.25~(1.43)\\
Cut 2 &0.31~(0.309)&0.28~(0.27)&0.19~(0.16)&&5.36~(5.89)&0.86~(0.97)&0.51~(0.58)\\
Cut 3 &0.26~(0.259)&0.25~(0.24)&0.17~(0.14)&&1.49~(1.64)&0.48~(0.54)&0.22~(0.25)
\\
\bottomrule[1.5pt]
\end{tabular}
 \end{table}

We present the cross sections of three typical signals ($m_T=1300, 1500, 2000$ GeV) and the relevant
backgrounds after imposing
the cuts in Table~\ref{cutflow1}, where the numbers in parentheses indicate the
results taking into account the ISR and beamstrahlung effects.
One can see that all the SM backgrounds are suppressed very efficiently, while the signals still have a relatively good efficiency at the end of the cut flow. The large background comes from the $e^{+}e^{-}\to W^{\pm}Zjj$ process, and the total cross section of SM backgrounds is about 0.002 fb.

\subsection{Analysis of  decay channels $Z\to \nu\bar{\nu}$ }
In this subsection, we analyze the signal and background events through the invisible decays $Z\to \nu\bar{\nu}$ decay channel.
 For this channel, the main SM backgrounds come from the following processes:
 \begin{itemize}
\item
$e^{+}e^{-}\to t\bar{t}$,
\item
$e^{+}e^{-}\to t\bar{t}Z$,
\item
$e^{+}e^{-}\to W^{\pm}Zjj$,
\item
$e^{+}e^{-}\to ZZjj$.
\end{itemize}

\begin{figure}[htb]
\begin{center}
\centerline{\hspace{2.0cm}\epsfxsize=10cm\epsffile{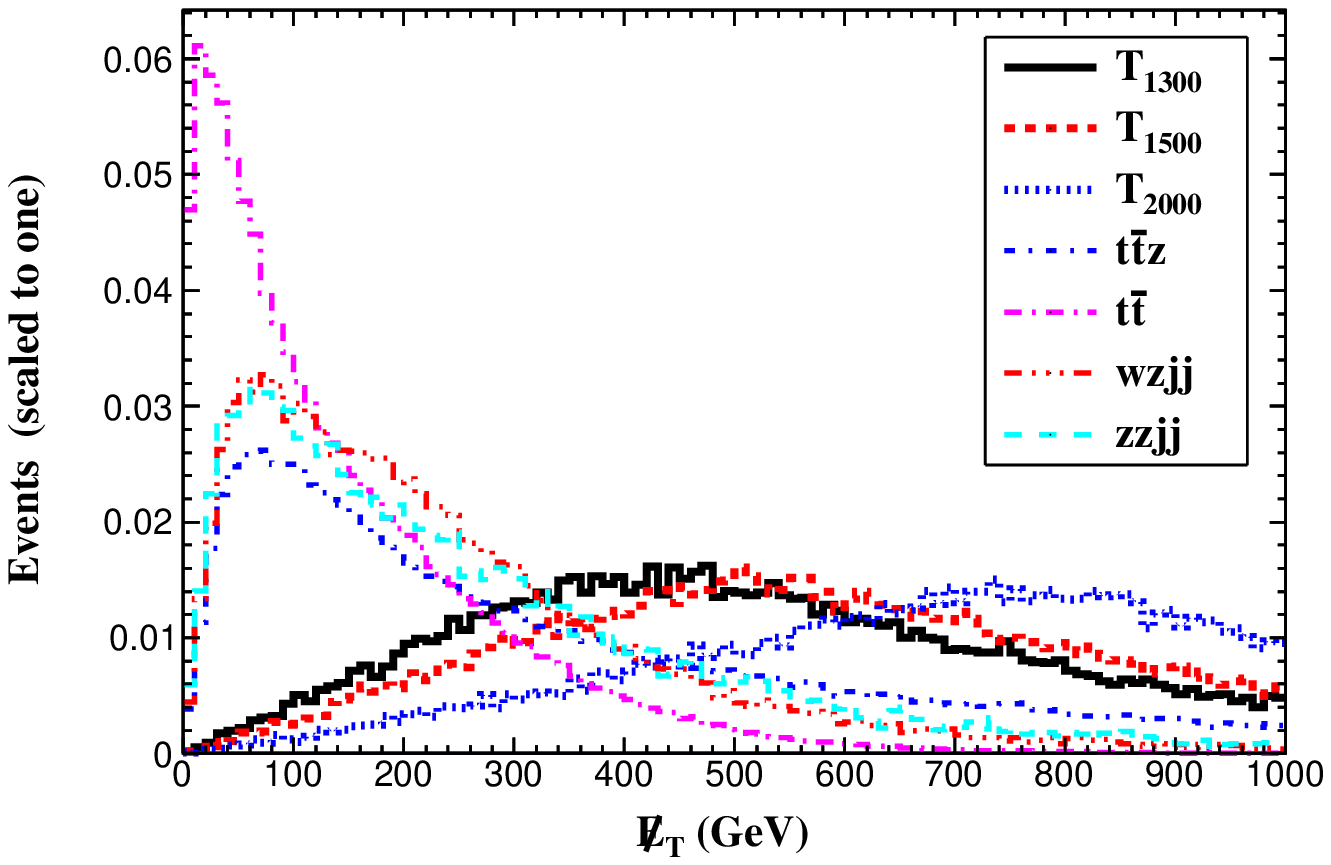}
\hspace{-2.0cm}\epsfxsize=10cm\epsffile{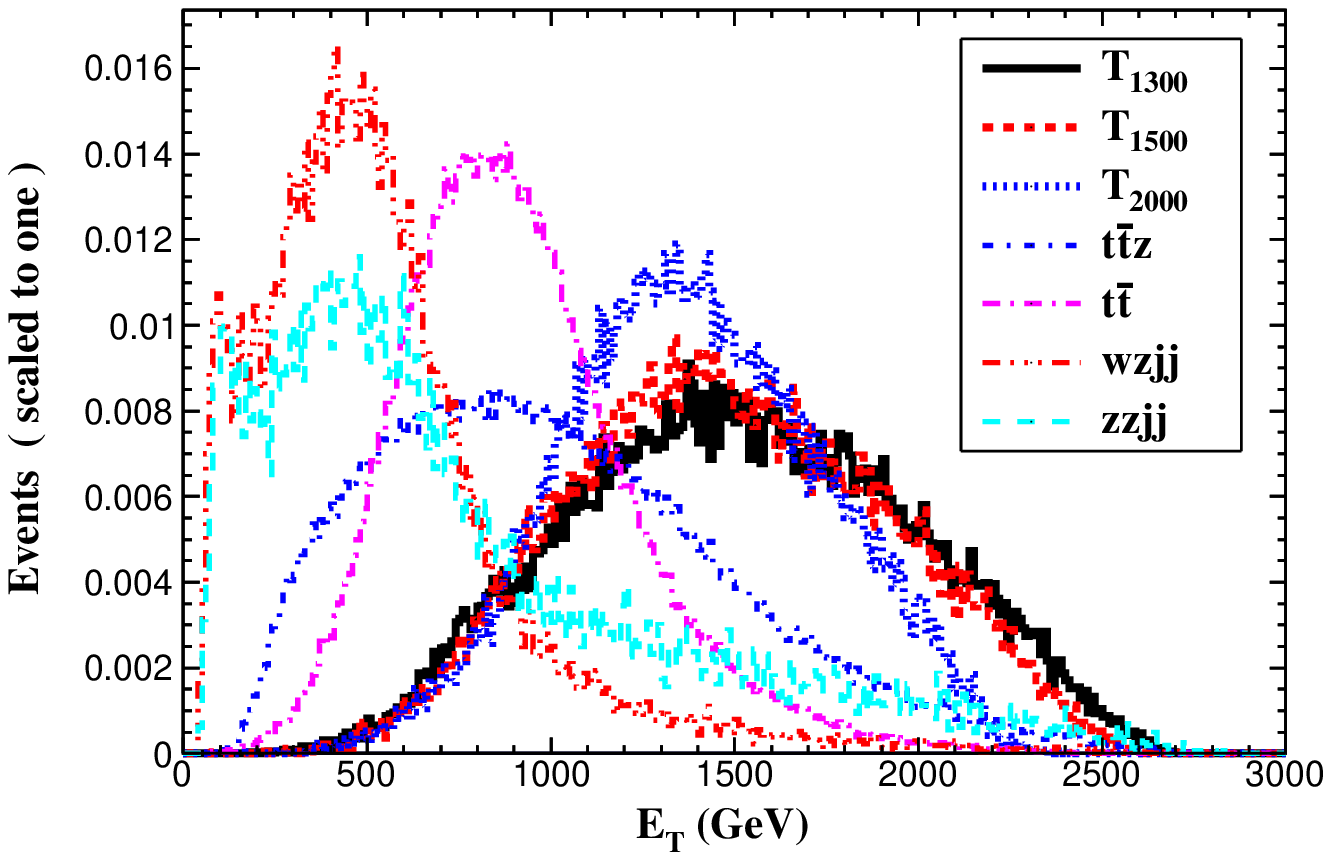}}
\caption{Normalized distributions for the signals (with $m_{T}$=1300, 1500 and 2000 GeV) and SM backgrounds at the CLIC. }
\label{distribution2}
\end{center}
\end{figure}

In order to get some hints of further cuts for reducing the SM backgrounds, we analyzed the normalized distributions of the missing transverse energy $\slashed E_{T}$,  and the scalar sum of the transverse energy of all final-state objects $E_{T}$ in Fig.~\ref{distribution2}.
Based on these kinematical distributions, a set of further cuts are given as follows:
 \begin{itemize}
\item Cut 1: Any electrons and muons are removed~($N(\ell)=0$),  at least three jets in which at least one is $b$ tagged. Furthermore, the transverse missing energy is required $\slashed E_{T}> 300 \rm ~GeV$.
\item Cut 2:
      The scalar sum of the transverse energy of all final-state objects $E_{T}$ is required to have $E_{T}> 1000 \rm ~GeV$. .
\item
Cut 3: The reconstructed $\slashed E_{T}$ is to be isolated from any jets by $\Delta\phi_{\slashed E_{T}j}> 0.8$.
\end{itemize}

\begin{table*}[htb]
\centering %
\caption{Same as Table~\ref{cutflow1}, but for the  decay channels $Z\to \nu\bar{\nu}$.\label{cutflow2}}
\vspace{0.2cm}
\begin{tabular}{p{1.6cm}<{\centering} p{2.0cm}<{\centering} p{1.8cm}<{\centering} p{1.8cm}<{\centering} p{0.3cm}<{\centering} p{1.6cm}<{\centering} p{2.0cm}<{\centering} p{2.0cm}<{\centering} p{1.8cm}<{\centering}}
\toprule[1.5pt]
 \multirow{2}{*}{Cuts}& \multicolumn{3}{c}{Signals}&\multicolumn{4}{c}{Backgrounds} \\ \cline{2-4}\cline{6-9}
&1300 GeV &1500 GeV& 2000 GeV  && $t\bar{t}Z$ &$t\bar{t}$&$WZjj$&$ZZjj$\\   \cline{1-9}  \midrule[1pt]
Basic&3.3~(3.293)&2.92~(2.77)&1.69~(1.40)&&133~(150)&3712~(5568)&1504~(1654)&86~(98)\\
Cut 1&1.57~(1.566)&1.45~(1.38)&0.92~(0.76)&&34~(39)&375~(562)&51~(56)&7.5~(8.55)\\
Cut 2 &1.24~(1.237)&1.18~(1.12)&0.76~(0.63)&&19~(22)&330~(495)&8.5~(9.4)&2.75~(3.14)\\
Cut 3 &0.77~(0.768)&0.74~(0.70)&0.48~(0.40)&&0.79~(0.89)&0.16~(0.24)&0.28~(0.31)&0.11~(0.13)
\\
\hline
\bottomrule[1.5pt]
\end{tabular}
 \end{table*}

We summarize the cross sections of three typical signals ($m_T=1300, 1500, 2000$ GeV) and the relevant
backgrounds after imposing
the cuts in Table~\ref{cutflow2}.
One can see that the $t\bar{t}$ background could be suppressed effectively by the cut $\Delta\phi_{\slashed E_{T}j}> 0.8$. The total cross section of the SM backgrounds is about  0.013 fb, which increases to 0.016 fb taking into account the ISR and beamstrahlung
effects.

\subsection{Discovery and exclusion significance }
In order to analyze the observability, we use the median significance  to estimate the expected discovery and exclusion significance~\cite{Cowan:2010js},
\be
\mathcal{Z}_\text{disc} &=
  \sqrt{2\left[(s+b)\ln\left(\frac{(s+b)(1+\delta^2 b)}{b+\delta^2 b(s+b)}\right) -
  \frac{1}{\delta^2 }\ln\left(1+\delta^2\frac{s}{1+\delta^2 b}\right)\right]} \\
   \mathcal{Z}_\text{excl} &=\sqrt{2\left[s-b\ln\left(\frac{b+s+x}{2b}\right)
  - \frac{1}{\delta^2 }\ln\left(\frac{b-s+x}{2b}\right)\right] -
  \left(b+s-x\right)\left(1+\frac{1}{\delta^2 b}\right)},
 \ee
with
 \be
 x=\sqrt{(s+b)^2- 4 \delta^2 s b^2/(1+\delta^2 b)}.
  \ee
Here, $s$ and $b$ are, respectively, the expected events of the signal and total SM background after all cuts, and $\delta$ is the percentage systematic  error.
In the limit of
$\delta \to 0$,  these expressions  can be simplified as
\be
 \mathcal{Z}_\text{disc} &= \sqrt{2[(s+b)\ln(1+s/b)-s]}, \\
 \mathcal{Z}_\text{excl} &= \sqrt{2[s-b\ln(1+s/b)]}.
\ee

\begin{figure}[htb]
\begin{center}
\vspace{-0.5cm}
\centerline{\epsfxsize=8cm \epsffile{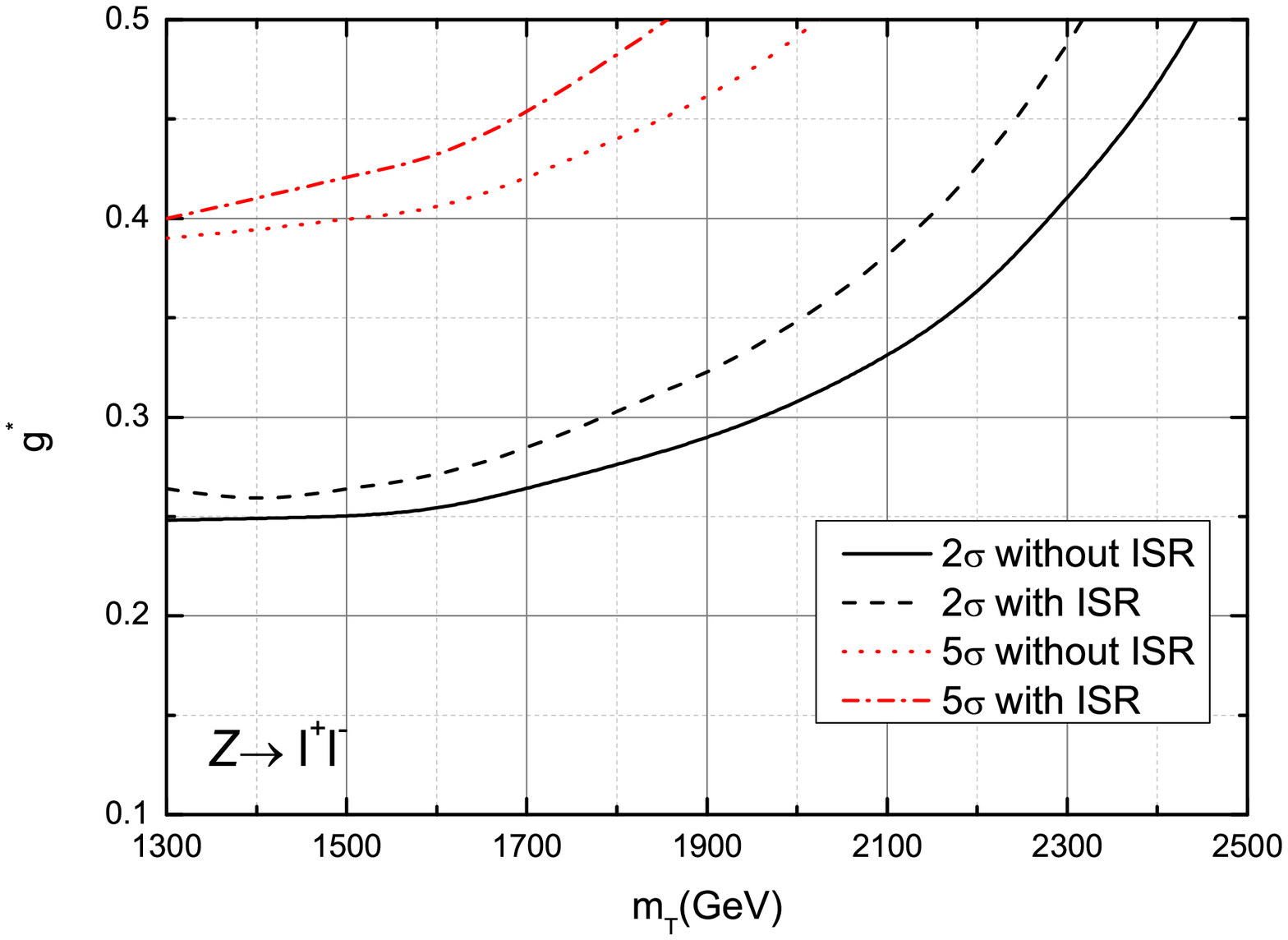}\epsfxsize=8cm \epsffile{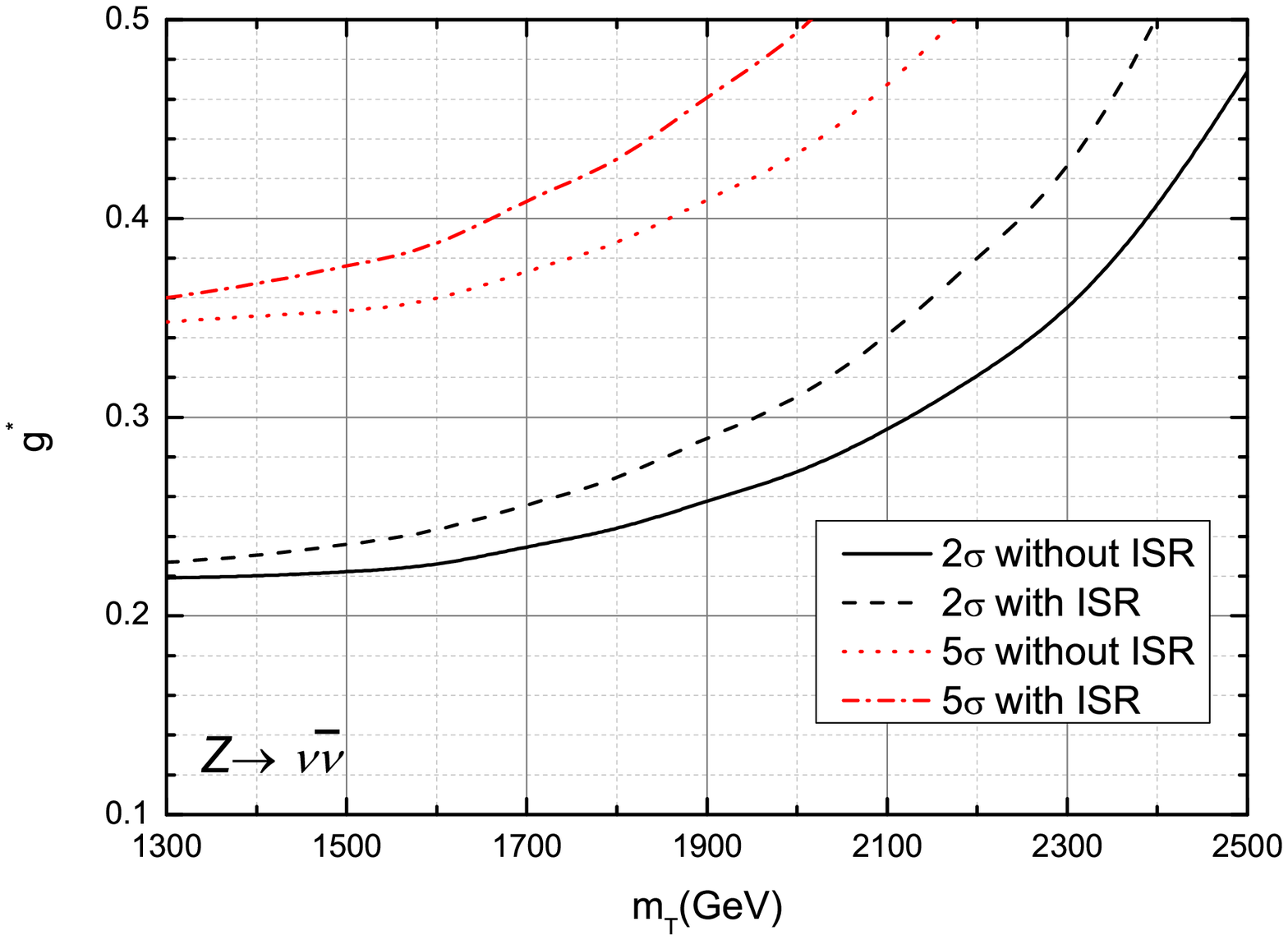}}
\caption{$2\sigma$ exclusion limit  and $5\sigma$ discovery prospects contour plots for $Z\to \ell^{+}\ell^{-}$ decay channel~(left), and for $Z\to \nu\bar{\nu}$ decay channel~(right) in $g^{\ast}-m_{T}$ planes at 3~TeV CLIC with integral luminosity 5 ab$^{-1}$. For simplicity, we here do not consider systematic uncertainties and take $\delta=0$. }
\label{ss}
\end{center}
\end{figure}

In Fig.~\ref{ss}, we plot the $2\sigma$ and $5\sigma$ sensitivity reaches with ISR and beamstrahlung effects and
without these effects, for the coupling strength  $g^{\ast}$ as a function of $m_T$ at 3~TeV CLIC with the integral luminosity 5 ab$^{-1}$. One finds that, for the $Z\to \ell^{+}\ell^{-}$ decay channel, the singlet VLQ-$T$ quark can be excluded in the regions of $g^{\ast}\in [0.25, 0.5]$ and $m_T\in$ [1300 GeV, 2450 GeV]  without the
ISR and beamstrahlung effects, while the discover regions can reach  $g^{\ast}\in [0.39, 0.5]$ and $m_T\in$ [1300 GeV, 2000 GeV].
While for the $Z\to \nu\bar{\nu}$ decay channel, the singlet VLQ-$T$ quark mass can be excluded in the regions of $g^{\ast}\in [0.22, 0.48]$ and $m_T\in$ [1300 GeV, 2500 GeV]  and the discover regions can reach  $g^{\ast}\in [0.35, 0.5]$ and $m_T\in$ [1300 GeV, 2180 GeV].
Besides, we can see that the ISR and beamstrahlung effects will decrease the excluding and discovering capability for the same VLQ-$T$ mass.

\begin{figure}[htb]
\begin{center}
\vspace{-0.5cm}
\centerline{\epsfxsize=8cm \epsffile{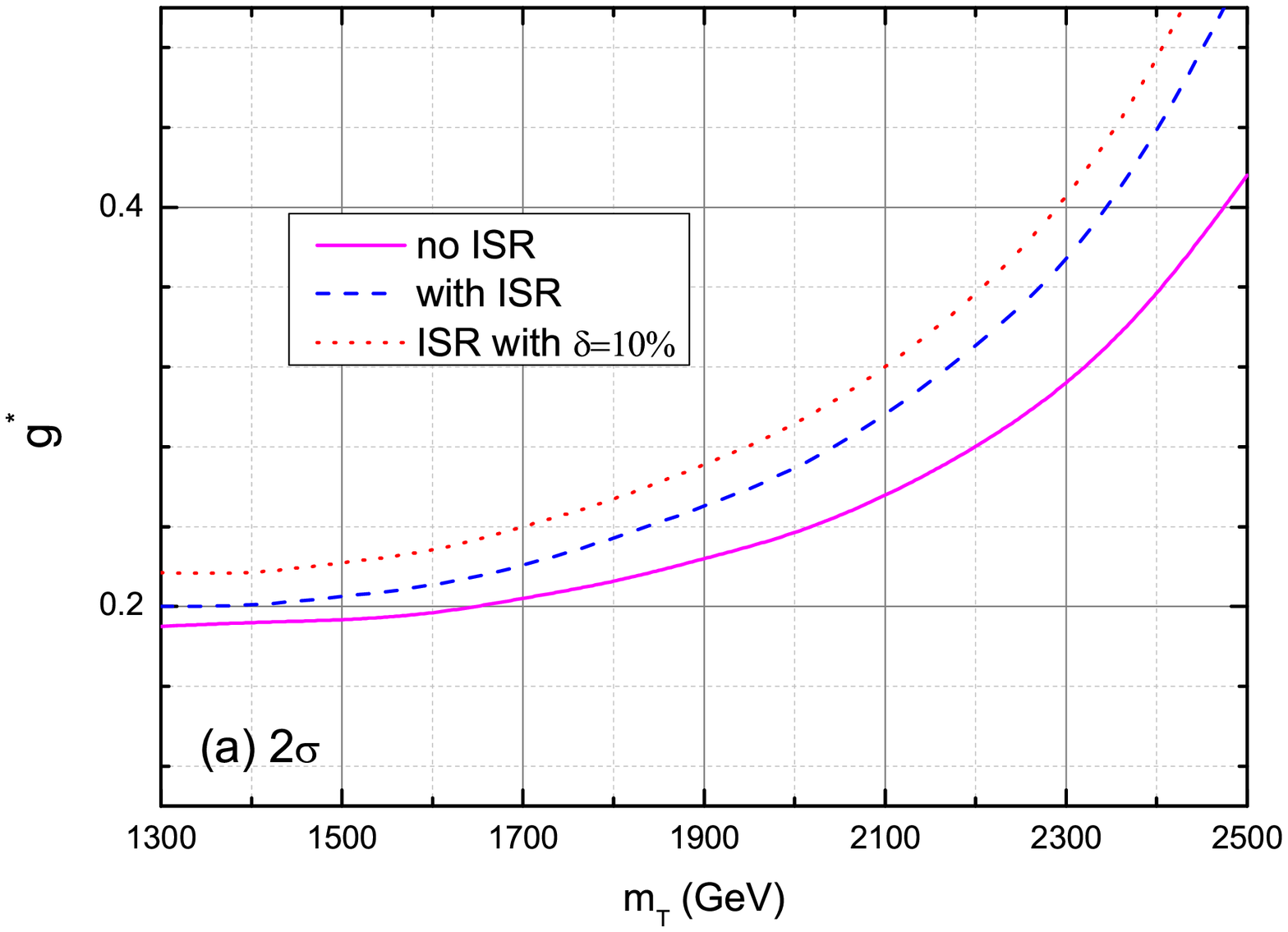}\epsfxsize=8cm \epsffile{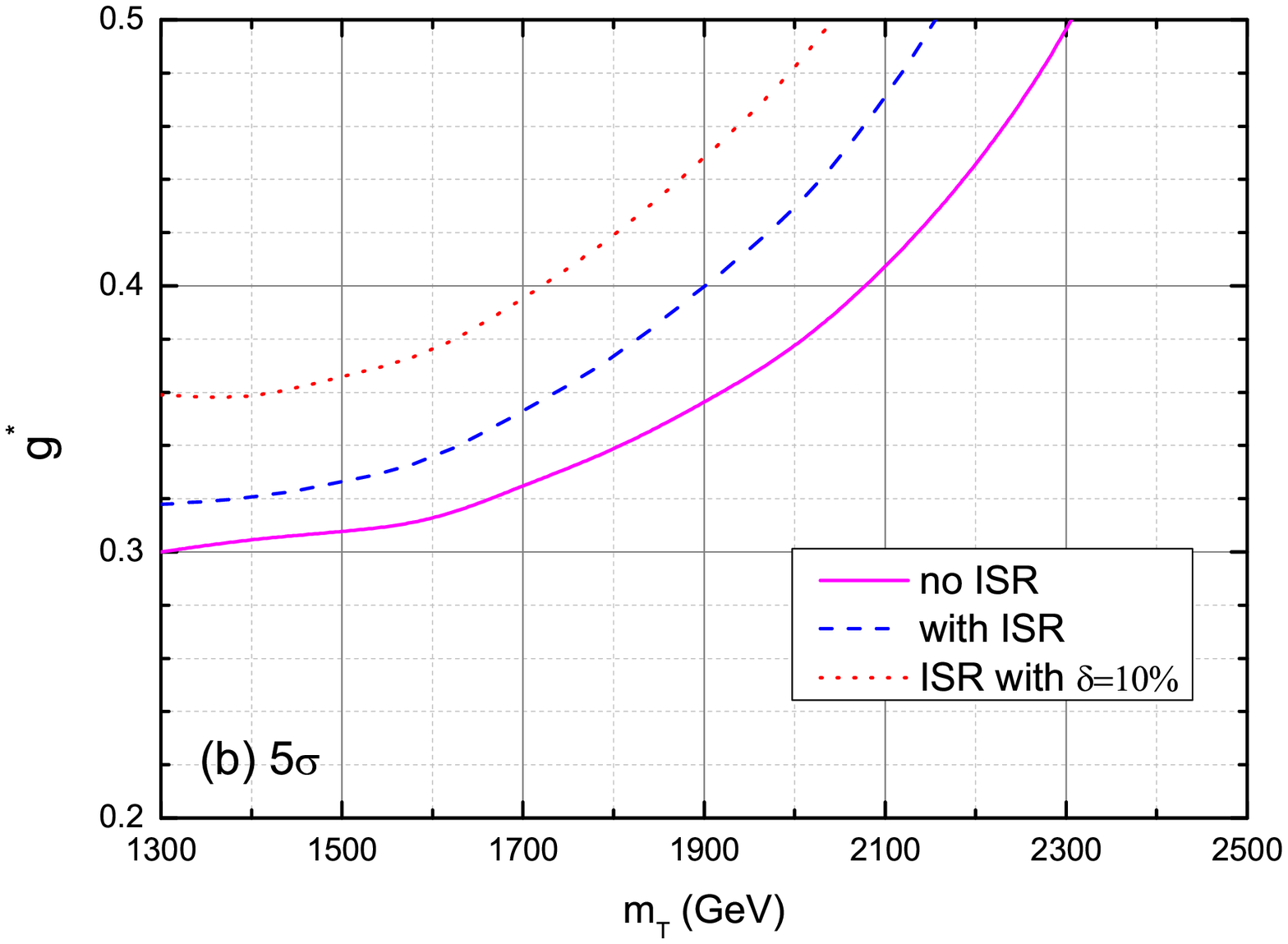}}
\caption{Combined $2\sigma$ exclusion limit and $5\sigma$ discovery prospect  contour plots for the singlet VLQ-$T$ signal in $g^{\ast}-m_{T}$ planes at CLIC with integral luminosity 5 ab$^{-1}$. }
\label{comb-s}
\end{center}
\end{figure}
\begin{figure}[htb]
\begin{center}
\vspace{-0.5cm}
\centerline{\epsfxsize=8cm \epsffile{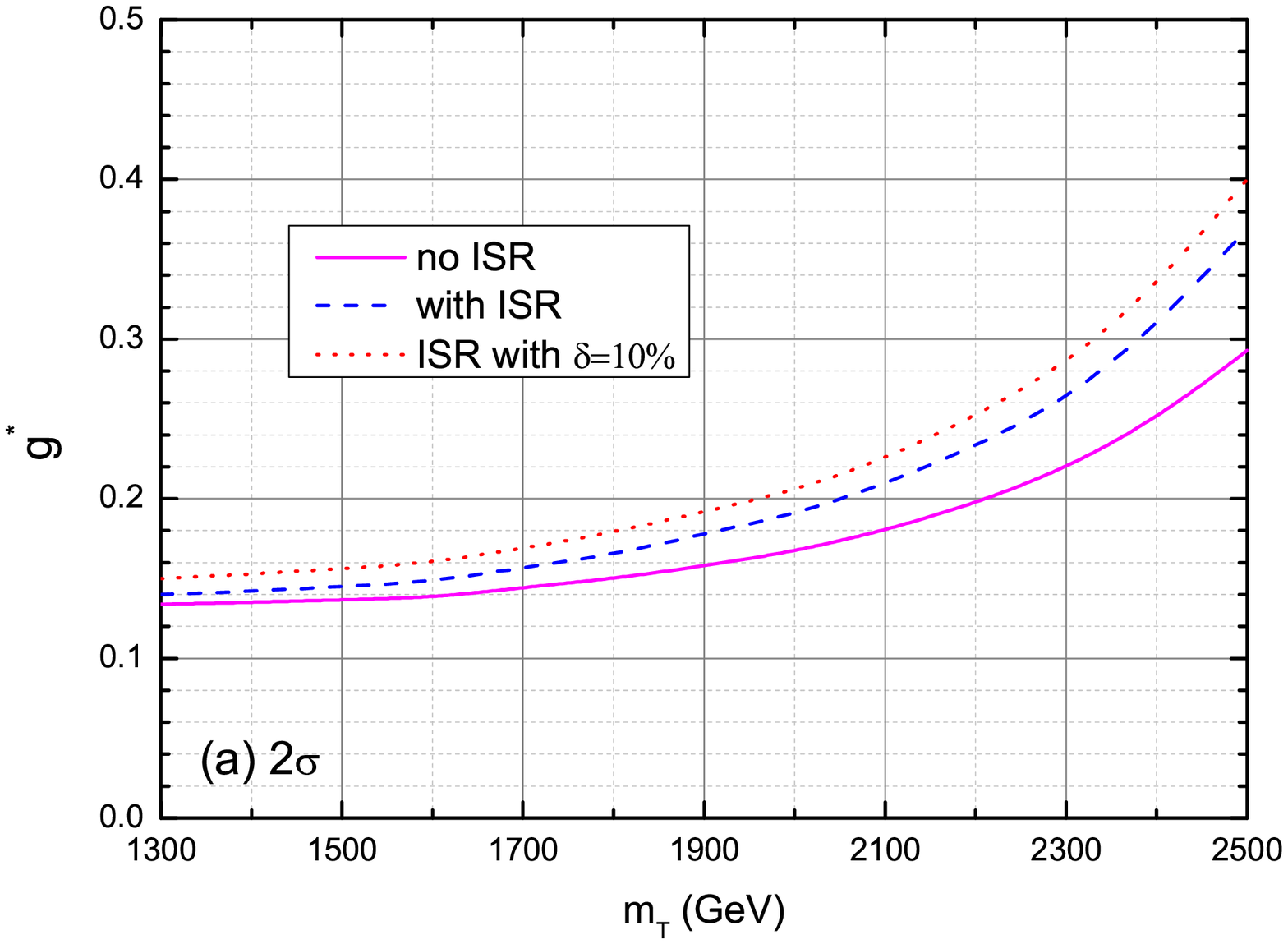}\epsfxsize=8cm \epsffile{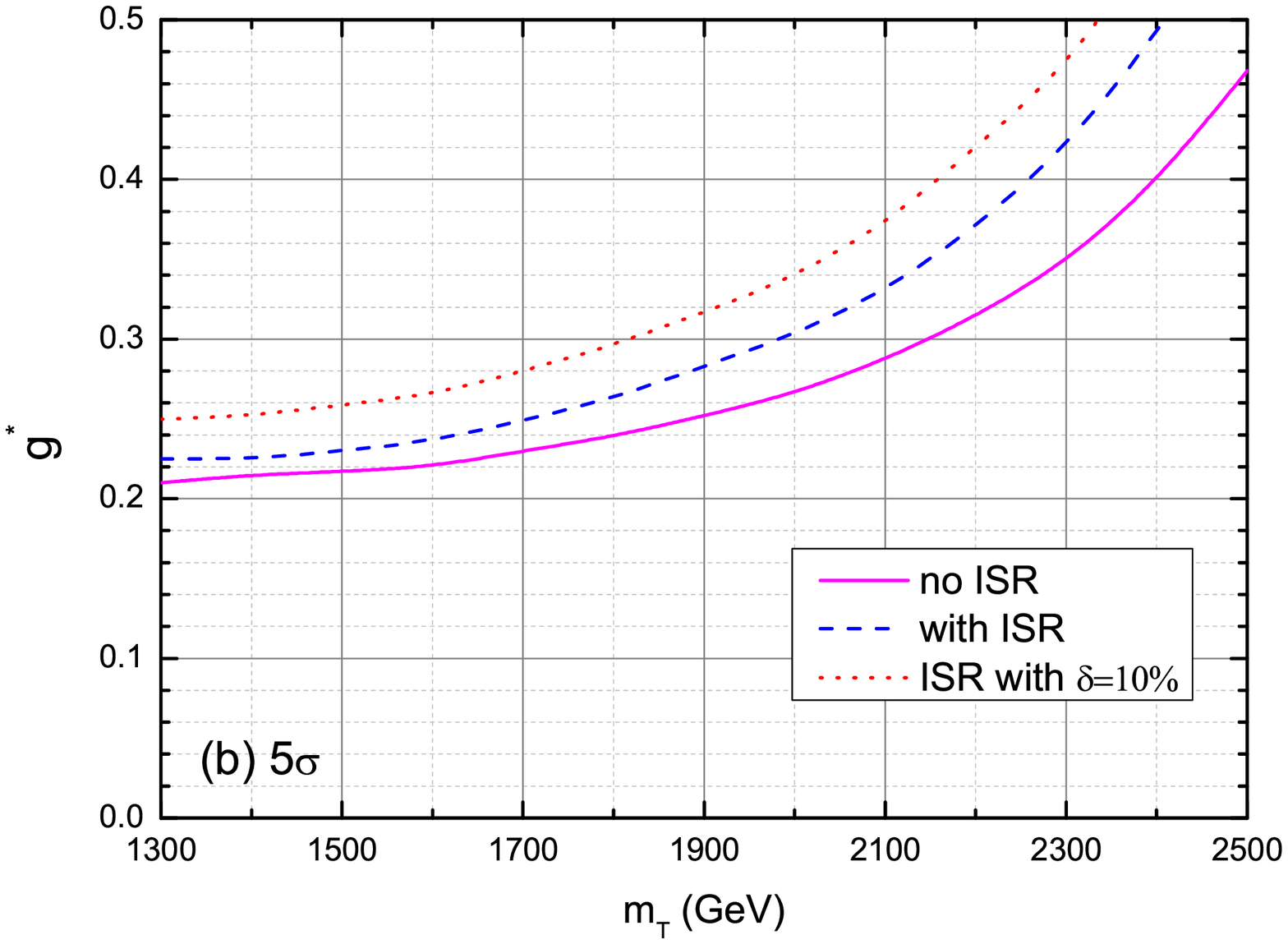}}
\caption{Same as Fig.~6, but for the doublet VLQ-$T$ case. }
\label{comb-d}
\end{center}
\end{figure}
Then, we combine its sensitivity with above two types of decay channels by using
$\mathcal{Z}_\text{comb}=\sqrt{\mathcal{Z}^{2}_{\ell\bar{\ell}}+\mathcal{Z}^{2}_{\nu\bar{\nu}} }$.
As we know, systematic uncertainty would weaken the excluding or discovery capability, so to illustrate the effect of systematic uncertainty on the significance, we will consider the case of  $\delta=10\%$.

For comparison, we further present in Figs.~\ref{comb-s} and~\ref{comb-d},  the combined  sensitivity reaches for the coupling strength  $g^{\ast}$ as a function of the singlet and doublet VLQ-$T$ quark mass $m_T$.
One can see that the singlet VLQ-$T$ can be excluded in the regions of $g^{\ast}\in [0.19, 0.4]$ and $m_T\in$ [1300 GeV, 2500 GeV]  at the 3 TeV CLIC with the integrated luminosity of 5 ab$^{-1}$, while the discover regions can reach $g^{\ast}\in [0.31, 0.5]$ and $m_T\in$ [1300 GeV, 2300 GeV].  For the doublet VLQ-$T$ case,  we found that the excluded correlation regions can be expanded to  $g^{\ast}\in [0.14, 0.3]$ and $m_T\in$ [1300 GeV, 2500 GeV], and the discovered
correlated regions can be expanded to  $g^{\ast}\in [0.21, 0.46]$ and $m_T\in$ [1300 GeV, 2500 GeV]. Besides, we
can see that the ISR and beamstrahlung effects and the systematic uncertainty will decrease  the
excluded and discovery regions.  For example, the excluded region
for the singlet VLQ-$T$ is decreased to $g^{\ast}\in [0.22, 0.5]$ and $m_T\in$ [1300 GeV, 2400 GeV], while the discovery regions are decreased to $g^{\ast}\in [0.36, 0.5]$ and $m_T\in$ [1300 GeV, 2000 GeV].
From the above discussions, we can see that, for single production of VLQ-$T$ with the subsequent $T\to tZ$ decay channel at 3 TeV CLIC, it is possible to detect its signal via above two decay channels.

Very recently, the ATLAS Collaboration has obtained the upper limit on the allowed coupling values $\kappa_T$ from a minimum value of
0.35 for $1.07 < m_T < 1.4 \rm ~TeV$ to 1.6 for $m_T = 2.3 \rm ~TeV$~\cite{ATLAS:2022ozf}.
Moreover, we list some
existing results related to searching for the singlet VLQ-$T$ in Table~\ref{list}. From this table, we can find that our result is competitive and complementary compared with
the previous studies.
Even though we worked in a simplified model including the singlet vectorlike top partner, our results are also mapped within the context of the specific models where the heavy $T$ quark only couples to the third generation
of SM quarks, such as the minimal composite Higgs model  with  singlet top quark partners.
From the couplings of the singlet top quark partner with the $W$ boson and a $b$ quark,
the mixing parameter $g^{\ast}$ is given by
$g^{\ast}\simeq \frac{y}{g}\frac{m_{W}}{m_{T}}$~\cite{Reuter:2014iya}, where $y$ is a Yukawa coupling controlling the mixing between the
composite and elementary states. For illustration, with $y=1$ and $m_T=1.5$~TeV, one obtains $g^{\ast}\simeq 0.16$.

\begin{table}[htbp]
\begin{center}
 \caption{\label{list}
Some results of searching for the singlet VLQ-$T$ at different colliders. ``$\setminus$" stands for no relevant results in the reference. }
\vspace{0.2cm}
\begin{tabular}{c|c|cc|cc|c}
\hline\hline
\multirow{2}{*}{Channel}      &\multirow{2}{*}{Data Set} &\multicolumn{2}{c|}{Excluding capability} & \multicolumn{2}{c|}{Discovery capability}&\multirow{2}{*}{Reference}  \\
\cline{3-4} \cline{5-6}
&&$g^{\ast}$&$m_T/\rm TeV$&$g^{\ast}$&$m_T/\rm TeV$&  \\
\hline
$T\to tZ$&LHC @14 TeV, 3 ab$^{-1}$&[0.06, 0.25] & [0.9, 1.5]&[0.10, 0.42]&[0.9, 1.5]&\cite{Liu:2017sdg}  \\
$T\to th$&LHC @14 TeV, 3 ab$^{-1}$&[0.16, 0.50] & [1.0, 1.6]&[0.24, 0.72]&[1.0, 1.6]&\cite{Liu:2019jgp}  \\
$T\to bW^{+}$&LHC @14 TeV, 3 ab$^{-1}$&[0.19, 0.50] & [1.3, 2.4]&[0.31, 0.50]&[1.3, 1.9]&\cite{Yang:2021btv}  \\
$T\to bW^{+}$&$e\gamma$ collider @2 TeV, 1 ab$^{-1}$&[0.13, 0.50] & [0.8, 1.6]&$\setminus$&$\backslash$&\cite{Yang:2018fcx}  \\
$T\to tZ$&$e\gamma$ collider @3 TeV, 3 ab$^{-1}$&[0.15, 0.23] & [1.3, 2.0]&[0.23, 0.50] &[1.3, 2.0]&\cite{Shang:2019zhh}  \\
$T\to th$&$e\gamma$ collider @3 TeV, 3 ab$^{-1}$&[0.14, 0.50] & [1.3, 2.0]&[0.27, 0.50] &[1.3, 2.0]&\cite{Shang:2020clm}  \\
$T\to bW^{+}$&$e^{+}e^{-}$ collider @3 TeV, 5 ab$^{-1}$&[0.15, 0.40] & [1.5, 2.6]&[0.24, 0.44] &[1.5, 2.4]&\cite{Qin:2022mru}  \\
 $T\to tZ$&$e^{+}e^{-}$ collider @3 TeV, 5 ab$^{-1}$&[0.19, 0.40] & [1.3, 2.5]&[0.31, 0.50] &[1.3, 2.3]&this work  \\
\hline
 \end{tabular}
 \end{center}
\end{table}

\section{Conclusion}
In this work, we have concentrated on the single production of the VLQ-$T$ at the future 3 TeV CLIC via the process $e^{+}e^{-}\to T\bar{t}$ in a simplified model, in which only two free parameters are included, the VLQ-$T$ mass $m_T$ and the EW coupling constant $g^{\ast}$. We have performed a full simulation for the signals and the relevant SM backgrounds based on the decay channel $T\to tZ$.
The $2\sigma$ exclusion limits and $5\sigma$ discovery prospects in the parameter planes of the two variables $m_T$ and $g^{\ast}$ have been  obtained at future 3~TeV CLIC, assuming 5 ab$^{-1}$ of  integral luminosity and unpolarized beams.

Our numerical results show that the singlet VLQ-$T$ quark can be excluded in the regions of $g^{\ast}\in [0.19, 0.4]$ and $m_T\in$ [1300 GeV, 2500 GeV],   while the discover regions can reach  $g^{\ast}\in [0.31, 0.5]$ and $m_T\in$ [1300 GeV, 2300 GeV].  For the doublet VLQ-$T$ case,   the excluded correlation regions can be expanded to  $g^{\ast}\in [0.14, 0.3]$ and $m_T\in$ [1300 GeV, 2500 GeV] due to the larger cross sections, while the discovered
correlated regions can be expanded to  $g^{\ast}\in [0.21, 0.46]$ and $m_T\in$ [1300 GeV, 2500 GeV].

 We also considered the ISR
and beamstrahlung effects and found that the excluding
or discovery capability would be reduced due to the decreasing cross sections of the signals. For the singlet VLQ-$T$, the excluded regions
 are decreased to $g^{\ast}\in [0.2, 0.5]$ and $m_T\in$ [1300 GeV, 2450 GeV], while the discovery regions are decreased to $g^{\ast}\in [0.32, 0.5]$ and $m_T\in$ [1300 GeV, 2150 GeV].
In addition,  the excluding or discovery capability would
be weakened if we considered the systematic uncertainty of
backgrounds.
Compared with the results of some previous phenomenological studies, we find that the future high-energy linear $e^{+}e^{-}$ collider will prove to be a promising hunting ground for such new particles which have electroweak strength interactions.

\begin{acknowledgments}
This work is supported by the key research and development program of Henan Province~(Grant No.~22A140019) and the Natural Science Foundation of Henan Province~(Grant No.~222300420443).
\end{acknowledgments}

\end{document}